\documentclass[twocolumn,prl,showpacs]{revtex4}
\usepackage{graphicx}
\usepackage{dcolumn}
\usepackage{bm}
\usepackage{color}

\begin{document}
\title{Finite-size scaling of correlation functions in \\one-dimensional Anderson-Hubbard model}
\author{Satoshi Nishimoto}
\affiliation{Leibniz-Institut f\"ur Festk\"orper- und Werkstoffforschung 
Dresden, D-01171 Dresden, Germany}
\author{Tomonori Shirakawa}
\affiliation{Department of Physics, Chiba University, Chiba 263-8522, Japan\\
Institut f\"ur Theoretische Physik, Leibniz Universit\"at Hannover, D-30167 Hannover, Germany
}

\date{\today}
\begin{abstract}
We study the one-dimensional Anderson-Hubbard model using the density-matrix 
renormalization group method. The influence of disorder on the 
Tomonaga-Luttinger-liquid behavior is quantitatively discussed. 
Based on the finite-size scaling analysis of density-density correlation 
functions, we find the following results: 
i) the charge exponent is significantly reduced by disorder at low filling and 
near half filling, 
ii) the localization length decays as $\xi \sim \Delta^{-2}$, where $\Delta$ 
is the disorder strength, independently of the on-site Coulomb interaction 
as well as band filling, and 
iii) the localization length is strongly suppressed by the on-site Coulomb 
interaction near half filling in association with the formation of the Mott 
plateaus.
\end{abstract}

\maketitle

We know that lattice disorder or defects in the electron system lead to 
various intriguing effects. One of the most famous examples is Anderson 
localization~\cite{Anderson58}; a large number of phenomena is explained 
in terms of the localization of quantum states~\cite{Kramer93}. 
Meanwhile, it is widely recognized that a full understanding 
of the experimentally observed phenomena has a need for taking correlation 
effects, as well as disorder, into account. Hence, the Anderson localization 
in the presence of electron-electron interactions has been an inevitable 
problem in the field of condensed matter physics~\cite{Lewenstein07}. 
A hot topic is the appearance of ``zero bias anomaly~\cite{Efros75}'', 
i.e., a suppression of the density of states $A(E)$ at the Fermi energy 
$E_{\rm F}$, by the interplay between disorder and short-range Coulomb 
interaction~\cite{Chiesa08,Song08}. Recently, it was reported that 
the similar anomaly occurs even in one-dimensional (1D) system~\cite{Shinaoka08}. 
It is possibly related to a vanishing of the photoemission spectral 
weight around $E_{\rm F}$ (or ``pseudogap-like'' behavior) observed 
in quasi-1D charge-transfer materials TTF-TCNQ and Bechgaad 
salts (TMTSF)$_2$X~\cite{Vescoli00,Sing03}.

Generally, 1D systems play a key role for elucidating the physical 
properties of solids because many features of the electronic states 
can be discussed rigorously. It has been also confirmed that the ground state and 
low-lying excitations can be described within the framework of the 
Tomonaga-Luttinger-liquid (TLL) theory~\cite{Solyom79} even in the 
presence of both disorder and electron-electron interaction~\cite{Loss92,Gogolin94}. 
Therefore, we are allowed to measure the effect of disorder as a modification 
of the TLL states. The properties of TLL are characterized by a few 
quantities; most notably, the charge exponent $K_\rho$ determines 
the long-range behavior of various correlation functions in the 1D 
metallic state. In this regard, a point to ponder is that the all 
eigenstates of a disordered 1D system are exponentially localized 
in the asymptotic sense~\cite{Abrahams79}.

In this Letter, we study a 1D Hubbard model in the presence of disorder, namely 
1D Anderson-Hubbard (AH) model. Using the density-matrix renormalization group 
(DMRG) technique~\cite{White92}, the system-size dependent density-density 
correlation functions are calculated for various on-site Coulomb interactions and 
disorder strengths. We then propose a finite-size scaling method of the correlation 
functions for obtaining the modified TLL charge exponent and localization length 
of the electrons. In consequence, a quadratic decay of the localization length 
with the inverse of disorder strength is confirmed for any interaction and 
band filling. We also find a drastic reduction of the charge exponent with 
disorder except around quarter filling and a strong suppression of the localization length with the on-site Coulomb interaction near half filling.

The Hamiltonian of the 1D AH model is written as
\begin{equation}
H = -t \sum_{i=1, \sigma}^{L-1} (c^\dagger_{i\sigma}c_{i+1\sigma} + H.c.) 
+ U \sum_{i=1}^{L} n_{i\uparrow}n_{i\downarrow}\nonumber
+ \Delta \sum_{i=1, \sigma}^{L} \varepsilon_i n_{i\sigma}
\label{hamiltonian}
\end{equation}
where $c_{i \sigma}$ is annihilation operator of an electron with spin 
$\sigma$ (=$\uparrow$ or $\downarrow$) at site $i$, $L$ is the system 
length, and $n_{i\sigma}=c_{i \sigma}^\dagger c_{i \sigma}$ is number 
operator. The nearest-neighbor hopping integral $t$ and on-site Coulomb 
interaction $U$ are assumed to be constant over the system. 
The random on-site potential $\varepsilon_i$ ($i=1, \cdots, L$) is defined 
by a box probability distribution ${\cal P}(\varepsilon_i)=\theta(1/2-|\varepsilon_i|)$ 
with the step function $\theta(x)$ and the disorder strength is 
controlled by $\Delta$. The band filling is $n=N/L$ where $N$ is 
the total number of electrons. We set $t=1$ as energy unit hereafter.

We consider the long-range behavior of the density-density correlation 
in the presence of disorder. We focus on the case of $U \ge 0$ and 
$0 < \Delta \lesssim U/2$, where the system is in the localized phase. 
Thus, the density-density correlation functions may be defined 
like~\cite{Saso85,Giamarchi88}
\begin{equation}
C(r)=e^{-\frac{\pi^2r}{6\xi}} C(r)|_{\Delta=0}
 \label{eqn:Cr1}
\end{equation}
with the asymptotic behavior in the absence of disorder
\begin{equation}
C(r)|_{\Delta=0}=-\frac{K_{\rho}}{(\pi r)^2}
+\frac{A\cos(2k_{\rm F}r)}{r^{1+K_{\rho}}}\ln^{-3/2}(r)+\cdots,
 \label{eqn:Cr2}
\end{equation}
where $\xi$ corresponds to the localization length of the electrons. 
Calculating the Fourier transformation of Eq.(\ref{eqn:Cr1}),
\begin{equation}
C(q_1)=\sum_{r=1}^L C(r) e^{iq_1r},
 \label{eqn:Cr3}
\end{equation}
with $q_1=2\pi/L$, we obtain 
\begin{equation}
C(q_1)=-\frac{K_\rho^\ast}{2\pi^2} \frac{e^{-\frac{\pi^2L}{6\xi}}-1}{e^\frac{\pi^2}{6\xi}-1}q_1^2
 \label{eqn:Cr4}
\end{equation}
for $q_1 \ll 1$ ($L \gg 1)$. 
The parameter $K_\rho^\ast$ is interpreted as a modified TLL charge exponent 
due to the disorder. In the limit of weak disorder,  $K_\rho^\ast$ is equivalent 
to $K_\rho$ and $\xi$ tends to be infinite. By the DMRG method~\cite{Ejima05}, 
we calculate $C(q_1)$ on random sampling $300$ ($500$) realizations 
of ${\cal P}(\varepsilon_i)$ for $L=128$, $112$, $96$, $80$, $64$, $48$, 
($32$, and $16$); then, we take an average of the results for 
each system size in order to obtain physically meaningful values 
of $C(q_1)$ [The averaged value is denoted as $\bar{C}(q_1)$]. 
For accurate calculation, the open-boundary conditions (OBC) are applied and 
we keep up to $m=2400$ density-matrix eigenstates in the renormalization procedure. 
Note that special attention should be paid to the convergence of the calculation 
because the DMRG wave function is apt to get trapped in a `false' ground state 
for disordered system. Thus, $K^\ast_\rho$ and $\xi$ will be estimated by fitting 
our numerical data of $\bar{C}(q_1)$ with Eq.(\ref{eqn:Cr4}).


\begin{figure}[t]
    \includegraphics[width= 6.0cm,clip]{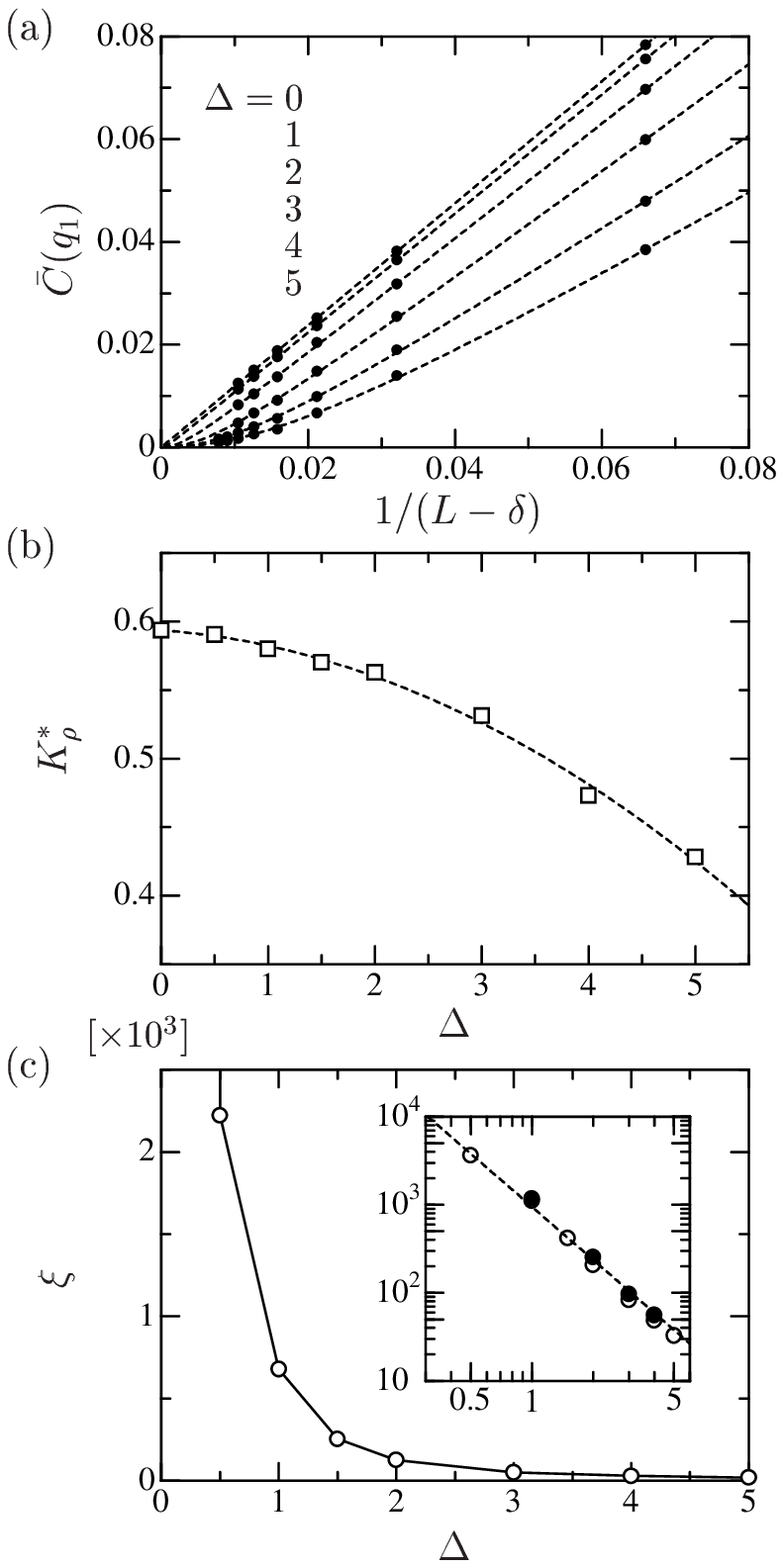}
  \caption{
(a) Finite-size-scaling analyses of the charge correlation function $\bar{C}(q_1)$ 
for various disorder strengths $\Delta$ at $U=10$. The dotted lines are 
the fitting with Eq.~(\ref{eqn:Cr4}). (b) Fitting results of the TLL charge 
exponent $K_\rho^\ast$ as a function of $\Delta$. The dashed line is 
a second-order polynomial in $\Delta$. (c) Fitting results of the localization 
length $\xi$ as a function of $\Delta$. Insets: Log-log plot of $\xi$ with $\Delta$ 
(open circles). The values of $\xi$ estimated from the Drude weight are also 
plotted as solid circles.
  }
    \label{fig1}
\end{figure}

For instance, we demonstrate the fitting of $\bar{C}(q_1)$ at $U=10$ and $n=1/2$. 
Figure~\ref{fig1}(a) shows the averaged values of $C(q_1)$ as a function 
of $1/(L-\delta)$ for various disorder strengths $\Delta$. Since a certain 
quantity of the correlations is missing at system edges due to the OBC, 
$L$ should be replaced with $L-\delta$ ($\delta>0$) for an excellent 
finite-size-scaling analysis. The correction factor $\delta$, which would 
be of the order of the lattice spacing, is determined to reproduce 
the relation $K_\rho^\ast=K_\rho$ in the absence of disorder; as expected, 
we obtain a quasi-infinite localization length ($1/|\xi| \lesssim 1.0 \times 10^{-5}$) 
with $\delta \approx 0.8-1.2$ at $\Delta=0$ for all fillings and interaction 
strengths considered in this study. In this way, the DMRG results of 
$\bar{C}(q_1)$ are well fitted by Eq.(\ref{eqn:Cr4}) even for finite $\Delta$ values, 
as shown in Fig.~\ref{fig1}(a). The estimated values of $K_\rho^\ast$ 
and $\xi$ are plotted as a function of $\Delta$ in Fig.~\ref{fig1} (b) and (c), 
respectively. The exponent $K_\rho^\ast$ is reduced with increasing $\Delta$ 
because the forward scattering processes are enhanced by the disorder~\cite{Giamarchi88}. 
However, the shortening of the localization length $\xi$ on $\Delta$ 
seems to be much more drastic. We find that $\xi$ decays as a power law 
with $\Delta$, i.e., $\xi=\xi_0 \Delta^{-\alpha}$, if $\Delta \lesssim U/2$. 
At $n=1/2$, we estimate $\xi_0 \approx 950$ and $\alpha \approx 2$ 
[see the inset of Fig.~\ref{fig1}(c)].

In order to check the accuracy of the above analysis, we also estimate 
the localization length from an exponential decay of the Drude weight 
$D(L)$ with the system length $L$~\cite{Kane92,Schmitteckert98} and perform 
a cross-check. The Drude weight is calculated on random sampling $100$ 
realizations of ${\cal P}(\varepsilon_i)$ for $64$, $48$, $32$, and $16$ 
using a recently proposed method~\cite{Shirakawa09}. The periodic 
boundary conditions are applied for this calculation. By fitting of 
the averaged results of $D(L)$ [$=\bar{D}(L)$] with a formula
\begin{equation}
\bar{D}(L)=\gamma \exp(-\frac{L}{\xi}),\ \ \ \ \ \ \gamma>0,
 \label{DL}
\end{equation}
we can obtain $\xi$ without any difficulty. As seen in the inset of Fig.~\ref{fig1}(c), 
there is an excellent agreement between the results from 
$\bar{D}(L)$ and $\bar{C}(q_1)$. Therefore, we can confirm 
the validity of our method.

\begin{figure}[t]
    \includegraphics[width= 5.5cm,clip]{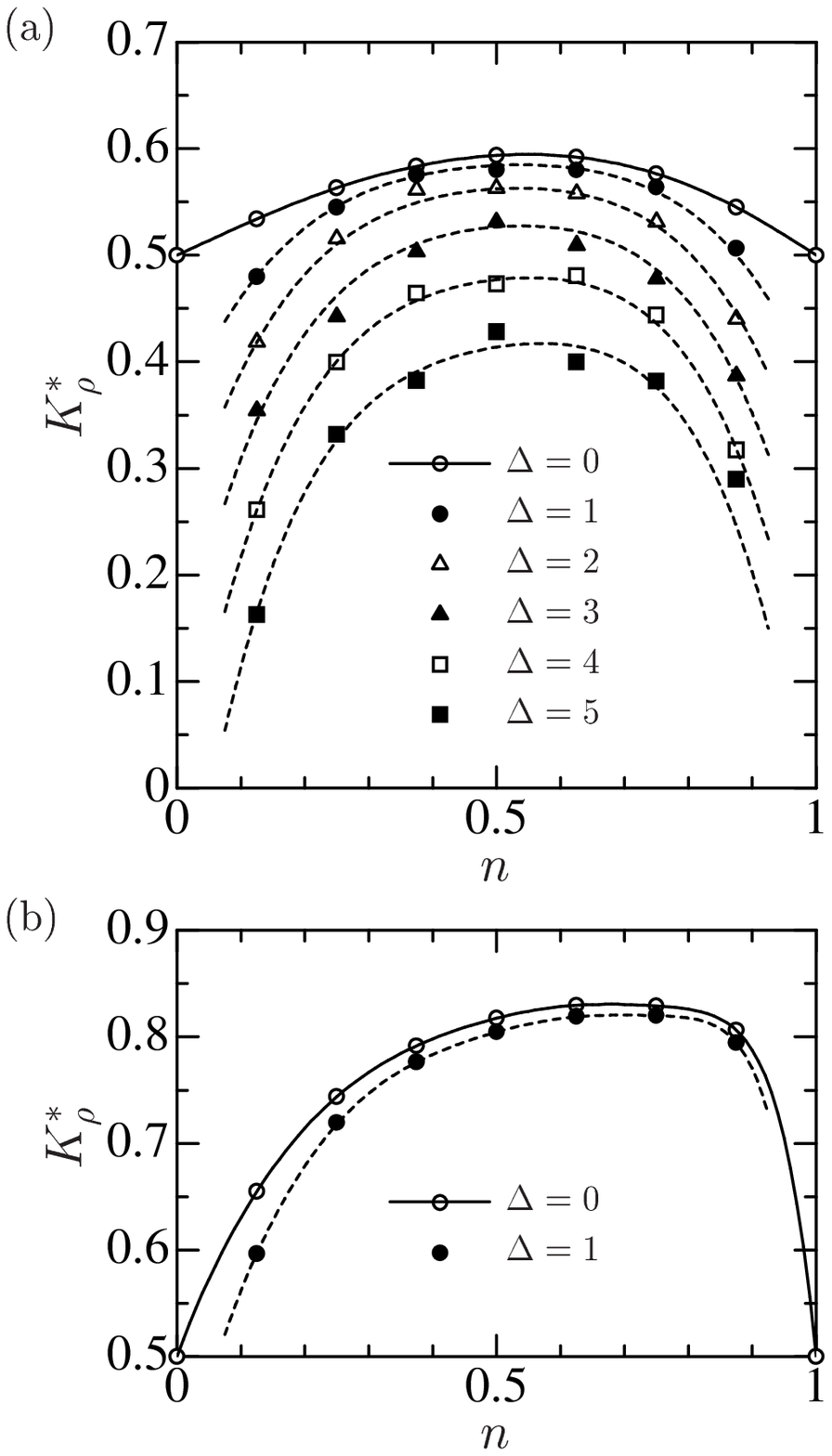}
  \caption{
Modified charge exponent $K_\rho^\ast$ as a function of the band filling 
$n$ for various values of the disorder strength $\Delta$ at (a) $U=10$ 
and (b) $U=2$. The dashed lines are guides to the eye.
  }
    \label{fig2}
\end{figure}

Let us now see how the charge exponent is modified by the disorder. 
Figure~\ref{fig2} shows the values of $K_\rho^\ast$ obtained by 
the above fitting procedure as a function of the band filling $n$ 
for various disorder strengths $\Delta$ at (a) strong ($U=10$) and 
(b) weak ($U=2$) interaction strengths. 
In the case of $U=10$, we see that the charge exponent is most severely 
affected by the disorder near $n=0$; namely, $K_\rho^\ast$ is drastically 
reduced by $\Delta$ since most of (or all) the electrons are easily trapped 
at sites having low on-site potential. Also near $n=1$, $K_\rho^\ast$ 
decreases rapidly with increasing $\Delta$. This can be explained 
in terms of the formation of Mott plateaus assisted by the disorder~\cite{Okumura08}. 
In other words, the system will be essentially in the Mott insulating 
state for tiny doping since the holes are strongly localized. 
On the other hand, the effect of disorder appears to be relatively 
weak around quarter filling because significant ``untrapped'' particles 
still remain and they could transfer. Consequently, the charge exponent is 
hardly affected for small $\Delta$ ($\lesssim 1$) at $n \approx 0.3-0.7$. 
It is consistent with a rather unstable Mott plateau far away from half 
filing~\cite{Okumura08}. We then turn to the case of $U=2$. Qualitatively 
the same behavior of $K_\rho^\ast$ with $\Delta$ is observed as long as 
the band filling is far away from half filing and, however, the effect of 
$\Delta$ near half filling seems to be weaker in comparison with 
the case of $U=10$. It would be concerned with the fact that the formation 
of the Mott plateau is harder for smaller $U$.

\begin{figure}[t]
    \includegraphics[width= 5.5cm,clip]{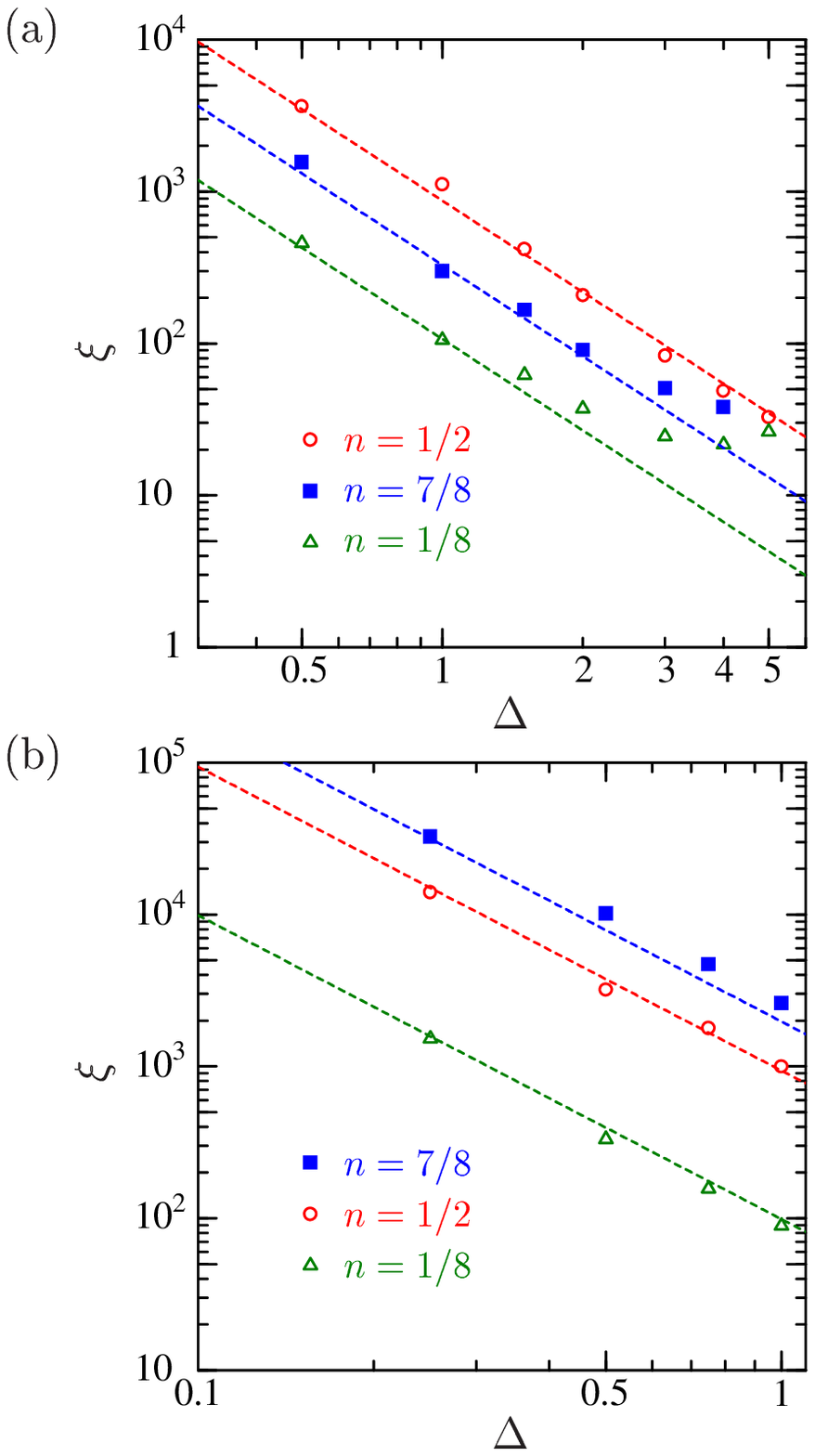}
  \caption{
The localization length $\xi$ as a function of the disorder strength $\Delta$ 
for (a) $U=10$ and (b) $U=2$. The dashed lines are the fitting function 
$\xi=\frac{\xi_0}{\Delta^2}$ for small $\Delta$.
  }
    \label{fig3}
\end{figure}

Of particular interest is the evolution of the localization length $\xi$
with increasing the disorder strength $\Delta$. Figure~\ref{fig3} shows 
the fitting values of $\xi$ as a function of $\Delta$ for several kinds 
of the band filling $n$ at (a) $U=10$ and (b) $U=2$. We find that 
the localization length always decay quadratically with the inverse of 
the disorder strength, i.e.,
\begin{equation}
\xi=\frac{\xi_0}{\Delta^2}
 \label{univ}
\end{equation}
when the disorder is weak. The same relation has been already proposed 
for spinless fermions case ($U=\infty$)~\cite{Thouless72,Herbert71}, 
which is the same model as what proposed by Anderson~\cite{Anderson58}. 
Therefore, the universality of Eq.(\ref{univ}) is confirmed in a weakly disordered 
system independently of the on-site Coulomb interaction and band filling. 
We note that the localization length begins to deviate from Eq.(\ref{univ}) 
at relatively small disorder $\Delta \approx 1.5-2$ near $n = 0, 1$ for $U=10$. 
It is because a part of the sites is doubly occupied under the strong 
disorder and the density-density correlation no longer obeys Eq.(\ref{eqn:Cr1}). 

\begin{figure}[t]
    \includegraphics[width= 5.5cm,clip]{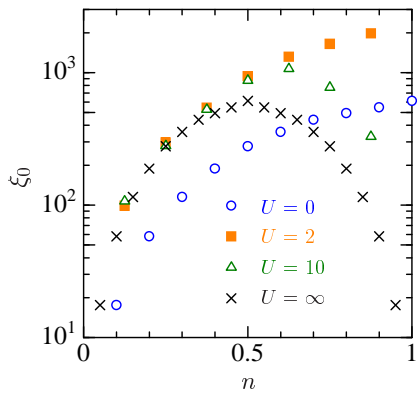}
  \caption{
Fitting values of the decay ratio $\xi_0$ for various interaction strengths.
  }
    \label{fig4}
\end{figure}

A qualitative determination of the decay ratio $\xi_0$ in Eq.(\ref{univ}) 
is another challenging problem.  So far, it has been studied 
perturbatively~\cite{Thouless72} and 
numerically~\cite{Czycholl81,Kappus81,Schmitteckert98} only for spinless fermions. 
Figure~\ref{fig4} shows the fitting values of $\xi_0$ as a function of the band 
filling $n$ at $U=0$, $2$, $10$, and $\infty$. When the interaction strength is 
varied from $U=0$ to $2$, $\xi_0$ increases for all fillings. It seems to be consistent 
with a prediction that the persistent currents are enhanced by the repulsive 
interactions~\cite{Giamarchi95} and, however, $\xi_0$ decreases with further 
increasing $U$. Especially near $n=1$, the reduction of $\xi_0$ with $U$ is 
quite rapid in connection to the stabilization of the Mott plateaus. 
On the other hand, surprisingly, $\xi_0$ is almost unchanged at $n \lesssim 0.5$ 
for $U \gtrsim 2$. Consequently, $\xi_0$ becomes symmetric about $n=1/2$ 
at large $U$, which reflects the particle-hole symmetry in the limit of 
$U \to \infty$. 

In summary, we consider the 1D Anderson-Hubbard model using the DMRG technique. 
A finite-size scaling method of the density-density correlation functions 
for obtaining the modified TLL charge exponent and localization length is 
demonstrated. As results, we find a quadratic decay of the localization length 
with the inverse of disorder strength for any interaction and band filling, 
a drastic reduction of the TLL charge exponent with disorder except around 
quarter filling and, a strong suppression of the localization length with 
the on-site Coulomb interaction near half filling.

Lastly, we make a short remark regarding an explanation of the pseudogap-like 
behavior and very small TLL exponent $K_\rho \approx 0.23$~\cite{Dressel96} 
observed in the TMTSF salts, which is quarter-filled system with dimerization. 
Regarding the dimer (TMTSF)$_2$ as a site, it may be reduced to a half-filled 
system. Thus, we will expect a strong reduction of the TLL exponent even 
by tiny disorder. Moreover, if the relation $A(E) \sim |E-E_{\rm F}|^\alpha$ 
with $\alpha=(K_\rho+K_\rho^{-1}-2)/4$ is still valid in the presence of the disorder, 
the strong reduction of $K_\rho$ would be compatible with the pseudogap-like 
behvbior.

\acknowledgments


\begin{thebibliography}{99}

\bibitem{Anderson58} P.W.~Anderson, Phys. Rev. {\bf 109}, 1492 (1958).
\bibitem{Kramer93} For a review, B.~Kramer and A.~MacKinnon, Rep. Prog. Phys. {\bf 56}, 1469 (1993).
\bibitem{Lewenstein07} For a review, M.~Lewenstein, A.~Sanpera, V.~Ahufinger, B.~Damski, 
A.~Sen, and U. Sen, Adv. Phys. {\bf 56}, 243 (2007).
\bibitem{Efros75} A.L.~Efros and B.I.~Shklovskii, J. Phys. C {\bf 8}, L49 (1975).
\bibitem{Chiesa08} S.~Chiesa, P..~Chakraborty, W.E.~Pickett, and R.T.~Scalettar, arXiv:0804.4463v1.
\bibitem{Song08} Y.~Song, S.~Bulut, R.~Wortis, W.A.~Atkinson, arXiv:0808.3356v1.
\bibitem{Shinaoka08} H.~Shinaoka and M.~Imada, \prl {\bf 102}, 016404 (2009).
\bibitem{Vescoli00} V.~Vescoli1, F.~Zwick, W.~Henderson, L.~Degiorgi, M.~Grioni, G.~Gruner, 
and L.K.~Montgomery, Eur. Phys. J. B {\bf 13}, 503 (2000).
\bibitem{Sing03} M.~Sing, U.~Schwingenschl\"ogl, R.~Claessen, M.~Dressel, and C.S.~Jacobsen, 
\prb {\bf 67}, 125402 (2003).
\bibitem{Loss92} D.~Loss, \prl {\bf 69}, 343 (1992).
\bibitem{Gogolin94} A.O.~Gogolin and N.V.~NV Prokof'ev, \prb {\bf 50}, 4921 (1994).
\bibitem{Solyom79} J.~S\'olyom, Adv. Phys. {\bf 28}, 201 (1979).
\bibitem{Abrahams79} E.~Abrahams, P.W.~Anderson, D.C.~Licciardello, and T.V.~Ramakrishnan, 
\prl {\bf 42}, 673 (1979).
\bibitem{White92} S.R.~White, \prl {\bf 69}, 2863 (1992); \prb {\bf 48}, 10345 (1993).
\bibitem{Saso85} T.~Saso, Y.~Suzumura, and H.~Fukuyama, Prog. Theor. Phys. Suppl. {\bf 84}, 269 (1985).
\bibitem{Giamarchi88} T.~Giamarchi and H.J.~Schulz, \prb {\bf 37}, 325 (1988).
\bibitem{Ejima05} S.~Ejima, F.~Gebhard, and S.~Nishimoto, Europhys. Lett. {\bf 70}, 492 (2005).
\bibitem{Kane92} C.L.~Kane and M.P.A. Fisher, \prl {\bf 68}, 1220 (1992).
\bibitem{Schmitteckert98} P.~Schmitteckert, T.~Schulze, C.~Schuster, P.~Schwab, 
and U.~Eckern, \prl {\bf 80}, 560 (1998).
\bibitem{Shirakawa09} T.~Shirakawa and E.~Jeckelmann, in preparation.
\bibitem{Okumura08} M.~Okumura, S.~Yamada, N.~Taniguchi, and M.~Machida, 
\prl {\bf 101}, 016407 (2008).
\bibitem{Thouless72} D.J.~Thouless, J. Phys. C {\bf 5}. 77 (1972).
\bibitem{Herbert71} D.C.~Herbert and R.~Jones, J. Phys. C {\bf 4}, 1145 (1971).
\bibitem{Czycholl81} G.~Czycholl, B.~Kramer, and A.~MacKinnon, Z. Phys. B {\bf 43}, 5 (1981).
\bibitem{Kappus81} M.~Kappus and F.J.~Wegner, Z. Phys. B {\bf 45}, 15 (1981).
\bibitem{Giamarchi95} T.~Giamarchi and B.S.~Shastry, \prb {\bf 51}, 10915 (1995).
\bibitem{Dressel96} M.~Dressel, A.~Schwartz, G.~Gr\"uner, and L.~Degiorgi, \prl {\bf 77}, 398 (1996).

\end{thebibliography}
\end{document}